\documentclass[twocolumn,showpacs,preprintnumbers,amsmath,amssymb]{revtex4}

\usepackage{graphicx}
\usepackage[latin1]{inputenc}
\usepackage{dcolumn}
\usepackage{bm}
\usepackage{epsfig}

\begin{document}

\preprint{APS/123-QED}

\title{Phase transition in the \emph{Countdown} problem}

\author{Lucas Lacasa$^{1,2,*}$ and Bartolo Luque$^1$}
\email{lucas.lacasa@upm.es}
\affiliation{$^1$Departamento de Matem\'atica Aplicada y Estad\'istica,
ETSI Aeron\'auticos\\ Universidad Polit\'ecnica de Madrid, Spain\\
$^2$ Department of Physics, Clarendon Laboratory, University of  Oxford, UK}%

\date{\today}

\begin{abstract}
Here we present a combinatorial decision problem, inspired by the celebrated quiz show called \emph{the countdown}, that involves the computation of a given target number $T$ from a set of $k$ randomly chosen integers along with a set of arithmetic operations.
We find that the probability of winning the game evidences a threshold phenomenon that can be understood in the terms of an algorithmic phase transition as a function of the set size $k$. Numerical simulations show that such probability sharply transitions from zero to one at some critical value of the control parameter, hence separating the algorithm's parameter space in different phases. We also find that the system is maximally efficient close to the critical point. We then derive analytical expressions that match the numerical results for finite size and permit us to extrapolate the behavior in the thermodynamic limit.
\end{abstract}

 \maketitle

In combinatorial optimization problems, a large amount of literature points out to the occurrence of a so called threshold phenomenon in the
performance of search algorithms \cite{selman, monasson, dimitri, moore}: there exists a phase in parameter space where the search algorithm can easily find a solution to the aforementioned
combinatorial problem (as the number of available solutions is exponentially large with the system size), and a phase where such solution typically (i.e. almost surely) does not exist. The transition between both phases is sharp in some situations, mimicking in several aspects the phenomenon of a phase transition in statistical physics problems. Some classical problems evidencing
such phenomenon include combinatorial problems in random graphs or the satisfaction of (random) boolean clauses, generically gathered under the umbrella
of random constraint satisfaction problems (rCSP) \cite{moore, bookSAT}. Many of these concrete problems can indeed by interpreted under a statistical physics formalism \cite{spin_glass, mertens2}, the general idea being the following: in a combinatorial optimization problem, in some cases one can formalize a cost function to be minimized. In satisfaction problems, this is for instance the number of violated constraints. Within statistical physics of disorders systems, such as in spin glasses, one indeed proceeds in the same manner if the system is studied in the limit of low temperature (in that situation, the system tries to adopt the ground state or minimal energy configuration). Thereby, the cost function of a combinatorial optimization problem can be related to the Hamiltonian of a disordered system at zero temperature (for instance, finding a minimum partition within the so-called partitioning problem is equivalent to finding the ground state of an infinite range Ising spin glass with Mattis-like, antiferromagnetic couplings \cite{mertens2}).\\
In this work we present a random decision problem, called the \emph{countdown} problem, which is inspired in a celebrated british TV quiz show called Countdown (based itself on the French game show \emph{Des chiffres et des lettres}, one of the longest-running game shows in the world, and receiving other names in several countries \cite{wiki}). This show is separated in several games, one of which incorporates the combinatorial problem of arithmetically combining some numbers to produce another one. Concretely, the contestants must use arithmetics to reach a given target number from six other numbers used each of them at much once. Here we formalize a random  version of this decision problem and explore its solvability as a function of the parameter space. We will provide numerical evidence according to which the solvability of the decision problem shows the presence of an algorithmic phase transition, and will introduce an approximate theory, which we show to be on good agreement with the numerics for finite size systems and permits us to theoretically find the phase transition in the thermodynamic limit.\\

\begin{figure}[h]
\leavevmode \epsfxsize=8.5 cm \epsffile{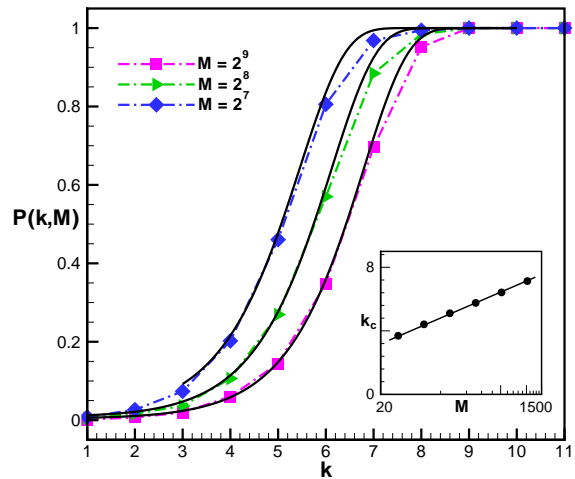}
\caption{Winning probability $P(k,M)$ curve as a function of $k$, for increasing pool sizes $M$, in the system with ${\cal A}=\{+,-\}$. Solid curves are the analytical predictions (equations \ref{P}-\ref{r(M)}). \emph{(Inset panel)} Linear-log plot of the crossover value scaling $k_c(M)$, for different values of the pool size $M$. In each case the crossover value is estimated as the linear interpolation of $k$ for which $P(k_c,M)=0.5$, as in percolation theory. The straight line is a fit to a logarithmic function $k_c(M)=a\log(M)+b$ with $a=0.98$ and $b=0.31$.}
\label{fig1}
\end{figure}

Let us begin by defining the \emph{pool} of size $M$ as the integer interval
$[1,M]$. Suppose that we randomly
extract with reposition from this pool a set of $k$ integers ${\cal X}=\{x_1, x_2, ..., x_k\}$ and
another integer $T$ called the target.
Stated as a decision problem, the target game raises the following question: for a given duple $(k,M)$, which is the probability $P(k,M)$
of reaching $T$ by combining the elements of $\cal X$ (where each element $x_i$ can be used or not, but each of them will be used at much once)
through the set of arithmetic operations ${\cal A}=\{+,-,\times,\div\}$? Once $M$ is fixed, one can
assert that for rather small values of $k$, as $k=2$ for example, the amount of possible
combinations is rather limited. For large values of $k$, the situation is the opposite: on
average there will exist many possible ways of combining the elements in ${\cal X}$ to reach $T$. Whereas several parallels with random k-satisfiability can be outlined, it can be shown that the problem at hands is not based on finding the correct variables assignment ${\cal X}$, but the correct Hamiltonian assignment ${\cal H}$ amongst a Hamiltonian ensemble \cite{futuro}. While this dual representation disables the possibility a standard statistical mechanics approach, we will take advantage of the number theoretical nature of the problem to propose a probabilistic theoretical treatment. We start be relaxing the problem statement by assuming that the set of available arithmetic operations is restricted to ${\cal A}=\{+,-\}$. As the problem is computationally nontrivial \cite{countdown}, in order to explore the problem numerically, we have implemented a brute force recursive routine that explores the search space in an exhaustive way (additional details in \cite{futuro}). For a given pool size $M$, we fix $k$, make Monte Carlo
 simulations and perform ensemble averages over different realizations of $T$ and
 $\cal X$. The winning probability $P(k,M)$ is defined as the probability of
 reaching $T$ by arithmetically combining at much the $k$ numbers (where each of the $k$ elements can be used or not, but each of them will be used at much once). In figure \ref{fig1} we plot the results of $P(k,M)$ vs $k$ for different pool sizes $M$, averaged in each case over $10^4$ realizations. Note that the transition
between $P\sim0$ (loosing almost surely) and $P\sim1$ (winning almost surely) is sharp and occurs at a certain $k_c(M)$, estimated as the linear interpolation of $k$ for which $P(k_c,M)=0.5$, as usual in percolation theory. This crossover value depends on the pool size, since the control parameter $k$ is not intensive. In the inset panel of figure \ref{fig1} we plot, in linear-log scales, its dependence with system's size,
finding a logarithmic scaling of the form $k_c(M)=a\log(M)+b$, with $a=0.98$ and $b=0.31$.

\begin{figure}[h]
\leavevmode \epsfxsize=8.5 cm \epsffile{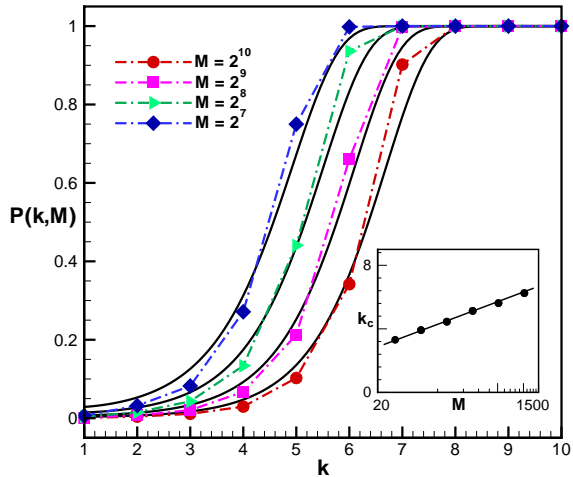}
\caption{Similar numerical results as figure \ref{fig1} in the general case of ${\cal A}=\{+,-, \times, \div\}$. The scaling of the critical point is again logarithmic $k_c(M)=a\log(M)+b$, with $a=0.84$ and $b=0.39$, what shifts the critical point towards smaller values of $k$. Solid lines are the results of the theory (equations \ref{P}-\ref{r(M)}) with the latter scaling relation.}
\label{fig2}
\end{figure}
\begin{figure}[h]
\leavevmode \epsfxsize=8.5 cm \epsffile{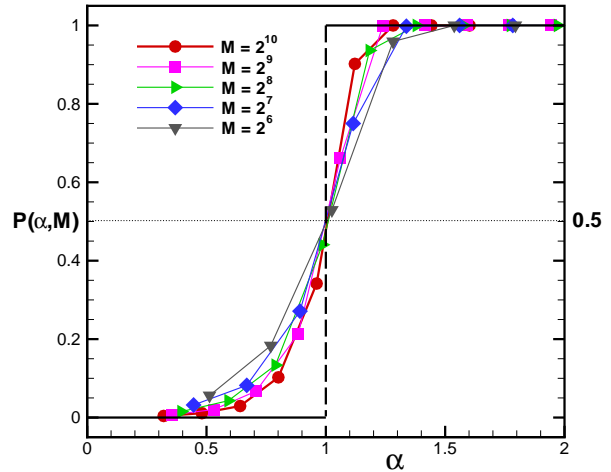}
\caption{Winning probability $P(\alpha,M)$ curve as a function of the intensive control parameter $\alpha=k/k_c(M)$, where $k_c(M)=a\log(M)+b$ is a scaling function of the crossover value with size (fitting values are $a=0.84$, $b=0.39$)
according to figure \ref{fig2}, for increasing system sizes. The curves get sharper as $M$ increases, pointing out to a phase transition in the thermodynamic limit: the onset of the so called threshold phenomenon. The Heaviside function, only reached in the thermodynamic limit, is a result of the theory (see the text).}
\label{collapsed}
\end{figure}

Our analytical treatment deals with the estimation of $P(k,M)$, when $k<<M$, and for that task we begin by considering a single operation, the sum. If we choose at random $n$ numbers from $[1,M]$, the largest possible value of its sum is $nM$. In a first approximation, we will suppose that the sum of $n$ numbers is uniformly distributed in $[1,nM]$. The result of this sum will fall in $[1,M]$ with probability $1/n$. The amount of results in $[1,M]$ accessible from $k$ numbers chosen at random from that interval is therefore $\sum_{n=1}^k{\frac{1}{n}\binom {k}{n}}$.\\
Let us proceed forward and introduce a second operation in our system: substraction. Incorporating this operation is approximately analogous to handle the former system, where now each number $a$ from the $n$ numbers chosen at random, represents both $a$ and $-a$.
 Now, the result of summing up the $2n$ numbers will belong to $[-nM,nM]$, such that, assuming uniformity once again, it will fall in $[1,M]$ with probability $1/2n$. Proceeding as before, the amount of results in $[1,M]$ accessible from $k$ numbers chosen at random from that interval, $N(k)$, can be written now as
\begin{equation}
N(k)=\sum_{n=1}^{2k}{\frac{1}{2n}\binom{2k}{n}}\approx \frac{2^{2k}-k-2}{2k+1}.
\label{N}
\end{equation}
The third approximation consists in assuming that those $N$ byproducts are indeed random independent trials of finding the target in $[1,M]$. In that situation, the winning probability $P(k,M)$ reads
\begin{equation}
P(k,M)= 1-\Big(1-\frac{1}{M}\Big)^{N(k)} \approx 1-e^{-\frac{N(k)}{M}},
\label{P}
\end{equation}
for $M>>1$. This result works qualitatively, however it can be improved, taking equation \ref{N} up to leading order and introducing dependency on $M$ through a modulating factor $r(M)$ that quantifies the correlations amongst numbers, such that
\begin{equation}
N(k,M) =\frac{e^{r(M) k}}{k}.
\label{ansatz}
\end{equation}
In order to estimate $r(M)$, note that $k_c(M)$ is defined such that $P(k_c, M)=1/2$. From equation \ref{P}, we find
\begin{equation}
r(M)=\frac{\log(M k_c \log 2)}{k_c}.
\label{r(M)}
\end{equation}
In figure \ref{fig1} we plot, in solid lines, the theoretical values of $P(k,M)$ as a result of equations \ref{P}-\ref{r(M)}, with the appropriate scaling $k_c(M)$ previously reported, which show good agreement with the numerics. In figure \ref{fig2} we extend the problem to the more general case where all elementary arithmetic operations are allowed, ${\cal A}=\{+,-,\times,\div\}$. The numerical results are analogous to the simpler case, where now the finite size scaling of the critical point fulfills $k_c(M)=a\log M+b$ with $a=0.84$ and $b=0.39$. Solid lines are the predictions of the theory with the latter scaling, showing again good agreement with the numerical simulations.



\begin{figure}[h]
\leavevmode \epsfxsize=8.5 cm \epsffile{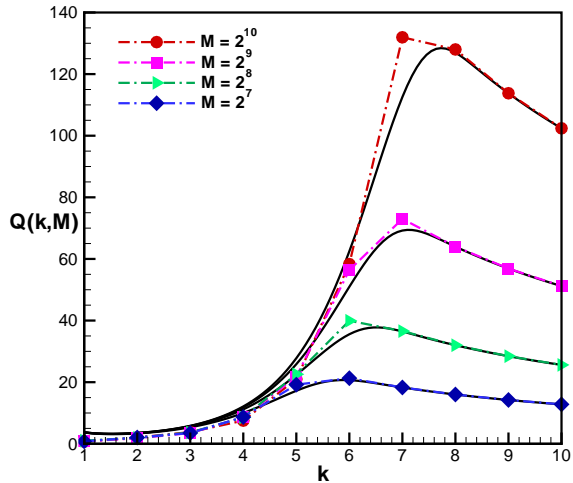}
\caption{Efficiency measure $Q(k,M)$ curve as a function of $k$, for increasing pool sizes $M$, in the system with ${\cal A}=\{+,-,\times,\div\}$. Solid curves are the analytical predictions (equation \ref{Q}). This measure is maximized near the transition point.}
\label{eficiencia}
\end{figure}

Introducing the intensive control parameter $\alpha=k/k_c$, in figure \ref{collapsed} we plot the numerical curves $P(\alpha,M)$ resulting from the simulations performed in the general case. As the system size increases, the probability of satisfying the game gets sharper around the now size independent crossover value. The behavior in the thermodynamic limit ($M\rightarrow\infty$, $\alpha$ finite) can be derived from equation \ref{P}-\ref{r(M)}, finding a Heaviside step function
\begin{equation}
P_{\infty}(\alpha) = \lim_{M \rightarrow \infty} 1-e^{-\frac{e^{r(M) \alpha k_c}}{\alpha k_c M}}=
\left\{
\begin{array}{rcl}
     0 & \textrm{if} & \alpha < 1
  \\ 1 & \textrm{if} & \alpha >1
\end{array}
\right.
\label{heaviside}
\end{equation}
i.e. the onset of a true phase transition.\\
Finally, let us define the function $Q(k,M)$ that measures the system's efficiency as the average amount of potential targets in $[1,M]$ that can be reached per unit number $k$. This function can be written as
\begin{equation}
Q(k,M)= \frac{P(k,M) M}{k},
\label{Q}
\end{equation}
whose behavior is shown in figure \ref{eficiencia}. This measure reaches a maximum in a neighborhood of $k_c(M)$, such that in the thermodynamic limit it diverges for $\alpha\rightarrow 1$, deepening on the conjecture that states that the complexity of multicomponent systems is maximized close to their critical points \cite{edge, edge2}.
 This behavior is also related to the so called easy-hard-easy pattern that takes place in algorithmic phase transitions, and suggests that close to the critical point (hard phase) the computation time that the algorithm needs to come along with a solution is maximized: for $\alpha<1$, the algorithm easily finds that the problem is unsatisfiable, whereas for $\alpha>1$ the algorithm easily finds one of many solutions, when $\alpha\sim1$ the number of solutions per unit number is optimal, and the algorithm needs to perform an exhaustive search of the whole space to find it. Note that within the quiz show Countdown, the pool is bounded to $M=1000$. Interestingly enough, the contestants are allowed to make use of $k=6$ numbers, what corresponds, according to our previous analysis, to the threshold between almost surely unsolvable to almost surely solvable instances. Driving the system towards the critical point assures that the game is hard but typically solvable, that is, interesting.\\

To summarize, in this work we have presented a random combinatorial decision problem that evidences an algorithmic phase transition separating the parameter space where the problem is either almost surely satisfiable or unsatisfiable. This work deepens on the relations between number
theory and theoretical physics, whose interface has shown to be a potential breeding ground \cite{number} for new approaches in both areas.
Some final remarks can be outlined: first, note that while in physical phase transitions the finite size scaling is usually in the form of a power law shape $k_c(M)\sim M^a$ with some finite size exponent $a$, in this problem we find logarithmic scaling $k_c(M)=a\log(M)+b$ (inset panels of figures \ref{fig1} and \ref{fig2}). From a thermodynamic viewpoint, the logarithmic scaling is not problematic, since it lacks an \emph{a priori} physical (i.e. thermodynamic) interpretation. If the reader is uneasy, we emphasize that in order to build a thermodynamic formalism we could always define an alternative control parameter $\tilde{k}\equiv\exp(k)$, in order to recover the usual power law scaling with system size $\tilde{k_c}(M)\sim M^a$, and make this parameter intensive (a temperature) through $\tilde{\alpha}=\tilde{k}/ M^a$. On the other hand, we note that logarithmic scalings have been found previously in other number-theoretic systems evidencing collective phenomena \cite{PRER,NJP,PRL}.

\noindent Finally, while the transition between unsatisfiable and satisfiable phases (loosing/winning) occurs at lower values in the general case ${\cal A} = \{+,-,\times,\div\}$ than in the simple one ${\cal A} = \{+,-\}$, in both situations the system evidences the threshold phenomenon. Is this phenomenon only restricted to elementary arithmetic systems or, much on the contrary, is this a fundamental behavior in abstract algebraic structures defined as a set of elements with some binary operations? On this respect, notice that the logarithmic scaling $k_c(M)$ can be interpreted as the growth rate of the minimal amount of elements needed to cover a growing system through binary operations, since the amount of possible outcomes of combined binary operations grows exponentially fast. This is a challenging open question for future research, which could be addressed within combinatorial group theory \cite{Gro93}.

\acknowledgments{}
We thank R\'{e}mi Monasson for pointing us out the relation of this problem to combinatorial group theory, and acknowledge financial support from
grants MODELICO, Comunidad de Madrid and FIS2009-13690. LL thanks the hospitality of the Systems and Signals group in the University
of Oxford, where part of this research was developed.

\bibliography{apssamp}

\end{document}